\documentclass[12pt]{iopart}
\usepackage{graphicx}
\usepackage[OT4]{fontenc}

\begin{document}

\title{Simple and Compact Nozzle Design for Laser Vaporization Sources}

\author{M. G. Kokish, M. R. Dietrich, B. C. Odom}
\address{Department of Physics and Astronomy, Northwestern University, Evanston, IL, USA.}

\ead{b-odom@northwestern.edu}

\begin{abstract}

We have developed and implemented a compact transparent nozzle for use in laser vaporization sources. This nozzle eliminates the need for an ablation aperture, allowing for a more intense molecular beam. We use this nozzle to prepare a molecular beam of aluminum monohydride (AlH) suitable for ion trap loading of AlH$^+$ via photoionization in ultra-high vacuum. We demonstrate stable AlH production over hour time scales using a liquid ablation target. The long-term stability, low heat load and fast ion production rate of this source are well-suited to molecular ion experiments employing destructive state readout schemes requiring frequent trap reloading. 

\end{abstract}

\section{Introduction}

The ability to trap and prepare molecular ions in a well-defined state can pave the way for achieving new quantum computing schemes \cite{Andre2006}, cold and state-controlled chemistry \cite{Willitsch2008} and precision tests of fundamental symmetries \cite{Schiller2005, DeMille2008, Leanhardt20111}. For trapped molecular ions, resonance-enhanced multiphoton dissociation (REMPD) is commonly used to perform state readout due to its high efficiency \cite{Staanum2010, Roth2006, Ni2014}. However, the destructive nature of the measurement requires frequent trap reloading, creating a unique set of criteria for the molecular ion source. The source should present a minimized gas load during pulses, to maximize the duty cycle of experiments requiring ultra-high vacuum (UHV) conditions. Using a source that presents a low heat load also minimizes outgassing degradation of the vacuum. Due to these stringent requirements, out of the many molecular ions of potential interest, a disproportionate number of labs have chosen to work with molecular ions that can be created simply from Doppler-cooled atomic-ion precursors \cite{Goeders2013, Wan2015}.

Loading of ions directly from a laser-ablated target can meet many of the above criteria \cite{Meng2013}, with extra steps sometimes required to purify the trapped ion sample. \cite{Chen2011}. However, this technique requires a suitable ablation target, and the large energies of ejected ablation products can present practical challenges when loading multiple species. A pure ionic sample can also be loaded by selectively photoionizing a neutral precursor formed using laser vaporization \cite{Ni2014}. In this scenario the laser vaporization can be done inside a nozzle that produces a molecular beam directed toward an ion trap \cite{Duncan2012}, effectively isolating the ionic byproducts from the ion trap. This method allows for production of a wider range of molecules, because the high density of the carrier gas inside the nozzle can be used as a reactant when forming ablation products in UHV.

We improve upon this type of laser vaporization source by implementing a cylindrical nozzle entirely composed of transparent fused silica, providing a number of advantages. First, implementations of laser vaporization nozzles have typically had an aperture that allows the ablation laser to reach the target, leading to depressurization of the nozzle and decreased molecular flux \cite{Duncan2012}. Using a transparent material for the nozzle removes the need for an aperture, preserving ablation products within the beam. Sealed nozzles have been used with similar intentions \cite{Jarrold1987,Neal2007}; however, the compact design of the fully transparent nozzle avoids large block assemblies consisting of a window and multiple channels. Second, the small inner volume of the nozzle reduces the overall gas load needed to achieve a molecular beam. Lastly, if the ablation target requires heating, the low thermal conductivity of fused silica reduces the undesirable heat flow leaving the target region, reducing vacuum degradation. Fused silica also outgasses more easily than commonly used nozzle materials such as stainless steel or alumina \cite{OHanlon2003}. Here we implement this nozzle design and demonstrate a stable source of aluminum monohydride (AlH) as a means to rapidly load the aluminum monohydride cation (AlH$^+$) into an ion trap.

Exploiting its highly diagonal Franck-Condon factors, AlH$^+$ was used to demonstrate fast rovibronic ground state cooling in an ion trap \cite{Lien2014}. AlH$^+$ was previously formed by exposing 10's of trapped Al$^+$ ions to increased background pressures. Conversion to AlH$^+$ occurred over the course of a few minutes, making trap loading the experimental rate limiting step by a couple orders of magnitude \cite{Seck2014}. To trap a pure sample of AlH$^+$ ions, one can instead selectively photoionize its neutral precursor AlH inside the trap. We use our nozzle to implement a molecular beam in order to deliver high densities of AlH to the ion trap. Laser ablation of aluminum in the presence of hydrogen gas leads to the formation of AlH and heavier aluminum hydrides \cite{Chertihin1993}. In order to provide a stable, long-lasting flux of AlH we prepare a liquid aluminum target \cite{Neal2007}. Using this nozzle design we demonstrate stable production of AlH with no degradation observed over hour time scales. We use a (2+1) resonance enhanced multiphoton ionization (REMPI) process in order to perform spectroscopy on AlH and form AlH$^+$ \cite{Zhang1988}. This source can be used to efficiently provide an ion trap over 100 AlH$^+$ ions in a few seconds.

\section{Experimental}
\subsection{Apparatus Overview}

\begin{figure}
\centering
\includegraphics[scale=0.45]{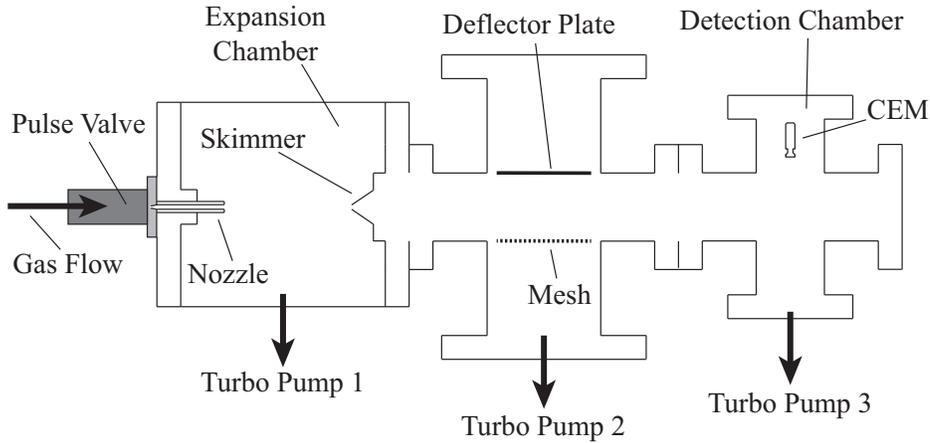}
\caption{\label{fig:chamb} AlH detection apparatus (not to scale). AlH emanating from the nozzle expands with the carrier gas. Ablation ions are extracted from the resulting molecular beam using a deflector plate. AlH is subsequently photoionized to form AlH$^+$, which is detected with a CEM.}
\end{figure}

The apparatus (Fig.~\ref{fig:chamb}) consists of two chambers: an expansion chamber and a six-way cross detection chamber separated by a four-way reducing cross that houses an ion deflector. The deflector, used to remove ionic ablation products, is realized by applying a high voltage ($\sim$500 V/cm over 3.5 cm) to a plate that extends as far as possible without completely obstructing the beam path. A mesh is placed on the opposite side in order to enhance the field near the molecular beam without obstructing the turbomolecular pump. The two chambers are separated by a skimmer (1 mm aperture, 20$^\circ$ inner, 25$^\circ$ outer half angle), which is mounted to a custom adapter plate mounted to a reducing flange. The chambers and the reducing cross are each pumped with a separate 50 l/sec turbomolecular pump \footnote{A separate turbomolecular pump for the reducing cross was not essential to AlH beam formation.}. The expansion chamber additionally consists of an ion gauge, a viewport on top and a nozzle. The nozzle is fitted to a custom flange consisting of copper feedthroughs and a ConFlat connection to a Parker Series 99 pulse valve. The pulse valve is driven at 200 V for 95 $\mu$s letting in H$_2$ with a backing pressure of 2 bar. The average pressure in the expansion chamber is 5x10$^{-4}$ and 1x10$^{-8}$ torr with the pulse valve activated and deactivated, respectively. The detection chamber has two viewports perpendicular to the molecular beam axis, which the photoionization beam passes through. The photoionization product is detected with a Photonis Magnum 5900 channel electron multiplier (CEM) perpendicular to the view ports and molecular beam. The CEM resides 30 cm downstream from the nozzle exit and its opening is situated 1 cm from the molecular beam axis.

\subsection{Nozzle}

\begin{figure}
\centering
\includegraphics[scale=0.45]{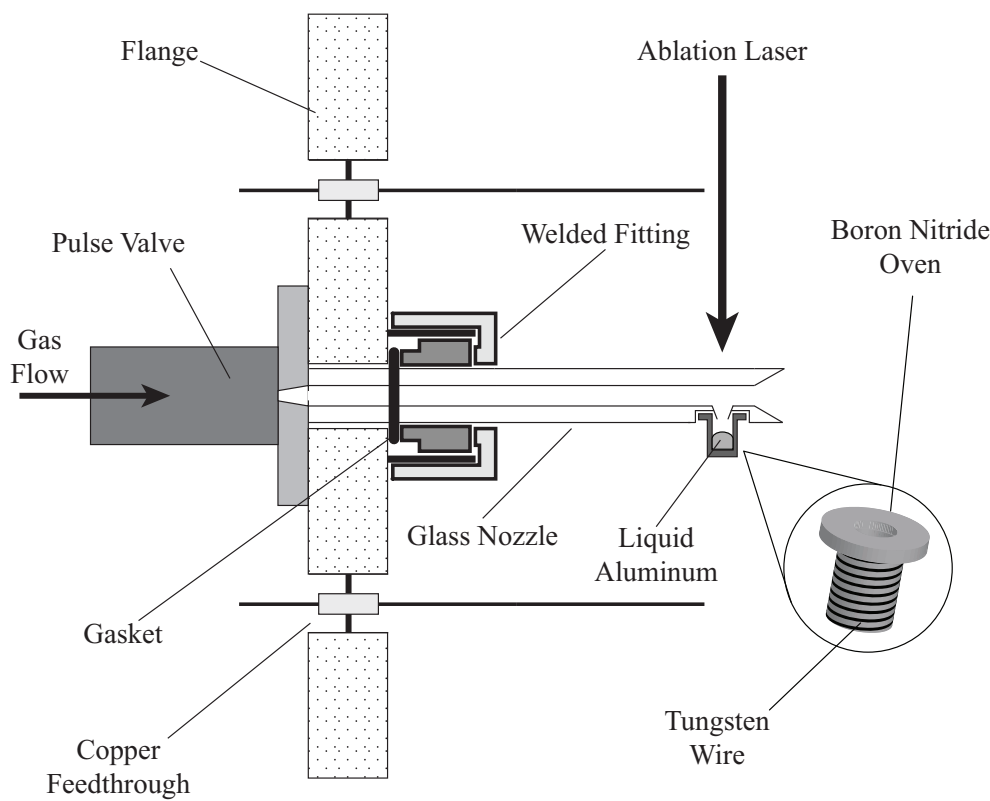}
\caption{\label{fig:nozz} Cross section of the nozzle assembly (not to scale). The clamp holding the boron nitride oven to the glass tube is not shown.}
\end{figure}

The nozzle (Fig.~\ref{fig:nozz}) is designed to overlap ablated liquid aluminum with a pulse of hydrogen. It consists of a 2.25 inch long fused silica tube (0.25 inch OD, 0.075 inch ID) that extends down to the pulse valve entrance and is held in place by a Swagelok Ultra-Torr fitting welded to the flange \cite{Zhu1992}. By using a transparent material, we avoid having to drill a hole for the ablation laser to pass through on its way to the sample target. Such a hole would depressurize the nozzle, leading to a decreased molecular flux. Using diamond-tipped drill bits in a water bath, we drill a 2 mm hole on the bottom side of the nozzle diameter that allows aluminum vaporized in the oven (described below) to enter the nozzle. The area around the opening is filed down such that the the nozzle can be seated inside the 4 mm opening of the oven, forming a press-fit seal. In addition, an 8 mm deep cone is drilled into the exit of the nozzle, such that the base of the cone is 3 mm from the oven exit hole. Allowing the beam to expand as soon as possible helps mitigate cluster formation \cite{Ebben1990}. The cone full angle is 30$^{\circ}$, which additionally helps collimate the beam \cite{Luria2011}.

The boron nitride oven (Kurt Lesker EVC10BN) holds solid pellets of 99.999\% aluminum. Boron nitride is inert to molten aluminum, ensuring that the aluminum does not move once melted. A saddle-shaped clamp with a hole in the center is draped over the nozzle, holding the oven tightly by grabbing onto the upper lip of the oven. This prevents the oven from moving during heating and helps seal any gaps between the oven and the glass. The hole in the clamp allows transmission of the ablation laser. Tungsten wire (0.375 mm diameter) is wrapped around the oven in order to provide heat. A snug fit is achieved by first wrapping the tungsten wire around a $\frac{1}{4}$-20 screw ($\sim$7-8 turns). This molds the tungsten wire into a coil having slightly smaller diameter than the 0.25 inch oven. The tungsten wire is connected to the copper feedthroughs using beryllium/copper inline barrel connectors. Melting of aluminum was observed at 6.5 A, 4 V; the tungsten wire glows red around the oven and yellow where the wire is not touching the oven. The ablation laser (532 nm, 10 Hz, 10 ns), is focused with an f = 150 mm lens. Initial measurements were performed with 11 mJ per pulse. Care must be taken not to ablate the walls of the glass tube as this causes the tube to crack,  preventing the ablation laser from reaching the target. 

\subsection{Detecting Photoions}

To find the photoionization signal we use an OPOTEK Magic Prism Optical Parametric Oscillator (OPO) pumped by the 3$^{\mathrm{rd}}$ harmonic (355 nm) generated from a Spectra Physics Quanta Ray Pro-270 Nd:YAG laser. The OPO is specified to emit a broad linewidth (20 cm$^{-1}$), which is expected to enhance the signal due to its ability to cover multiple rotational transitions. For these procedures we measured approximately 10 mJ of REMPI light entering the chamber. The light is focused to the molecular beam axis with an f = 200 mm lens and we use a photon counter (SRS SR400) to count individual ion events from the CEM. Because the AlH product yield is lower than that of aluminum, we first measured aluminum photoions. Aluminum has two strong (2+1) REMPI lines at 445.2 nm and 446.3 nm \cite{Mitchell1983}, conveniently close to AlH's Q branch transition centered at 448.5 nm \cite{Zhang1988}. 

\begin{figure}
\centering
\includegraphics[scale=0.4]{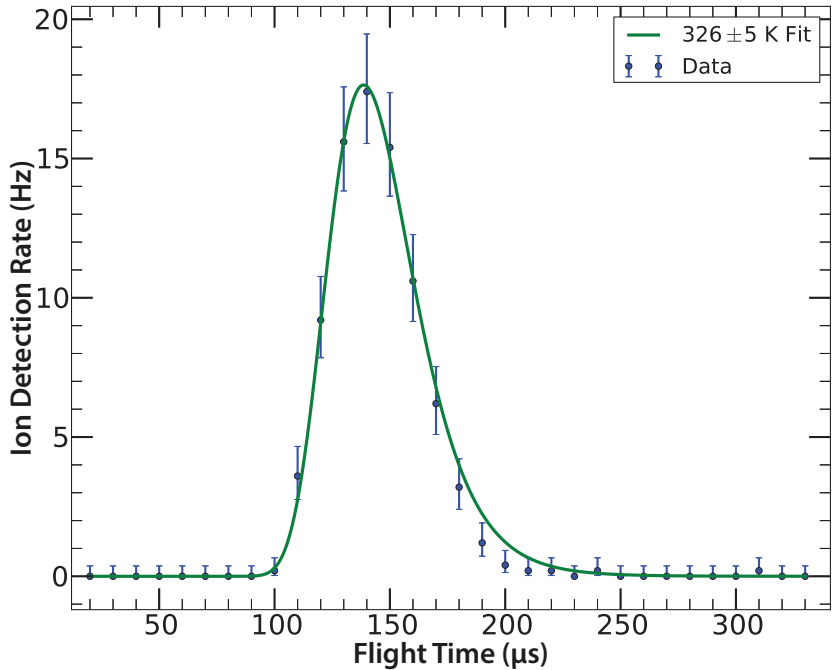}
\caption{\label{fig:tof} Time-of-flight profile for AlH with temperature fit from \cite{Christen2006}. Error bars correspond to 1$\sigma$ confidence intervals for a Poisson process. }
\end{figure}

Note that the nozzle acts as a cylindrical lens and further focuses the ablation laser, leading to higher intensities. Effective ablation of the aluminum target can be confirmed by visually observing a ``spark'' where the ablation laser contacts the target \cite{Duncan2012}. We observe that the highest photoionization signal occurs with the REMPI beam focused directly under the CEM and centered on the molecular beam's axis. As is common for laser vaporization sources, we find the signal to be very sensitive to both the REMPI and ablation pulse delays relative to the pulse valve trigger \cite{Tokunaga2007}. Ablating the molten aluminum target also leads to a background signal of ionic ablation products. However these ablation ions can be distinguished from REMPI ions due to their different arrival times at the CEM (10's of microseconds versus 150 $\mu$s) and are also mitigated by the ion deflector. A photodiode detects the arrival of the REMPI laser pulse in order to more easily facilitate synchronous detection of REMPI products. The REMPI ions are found to arrive at the CEM 700 ns after the REMPI pulse.

Due to the similarity in mass, Al and AlH have very similar velocity profiles. Switching to the Q branch of AlH (448.5 nm) led to a smaller photoionization signal. After rescanning the REMPI (Fig.~\ref{fig:tof}) and ablation pulse delay times, the AlH signal was then further optimized by increasing to higher hydrogen backing pressures and adjusting the ablation pulse intensity. We observe a signal saturation with increasing ablation pulse energy (2 mJ), and at higher intensities droplets of aluminum eject from the oven. We find that these droplets react with the glass, irreversibly altering its surface. We have also used deuterium as the carrier gas and confirm AlD's Q-branch peak at 448.1 nm, deduced using spectroscopic constants from \cite{Zhu1992}.

\section{Results}
\subsection{AlH Spectroscopy}

\begin{figure}
\centering
\includegraphics[scale=0.3]{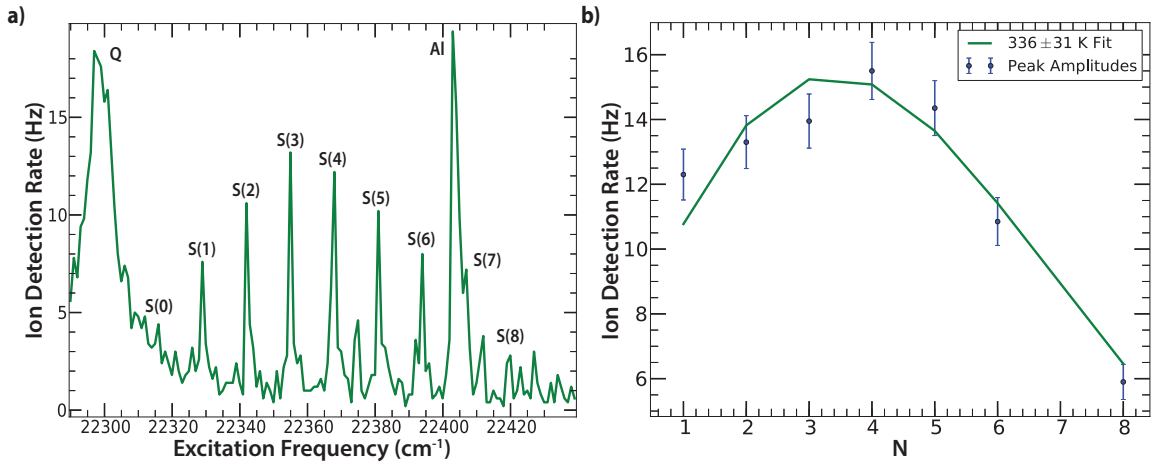}
\caption{\label{fig:spec}a) REMPI spectrum of the molecular beam. b) Rotational temperature fit to peak amplitudes of different S-branch transitions. S(0) and S(7) are omitted due to spectral interference. Error bars correspond to 1$\sigma$ confidence intervals for a Poisson process.}
\end{figure}

After optimizing the signal, we then switch to a narrow linewidth pulsed laser (0.06 cm$^{-1}$, Sirah PrecisionScan, Coumarin 450) and perform spectroscopy. Fig.~\ref{fig:spec}a shows a spectrum taken with 2 mJ of REMPI light focused with an f = 200 mm lens, integrating 5 seconds per point. The Q and S branches are easily identified. The spacing of the S-branch peaks yields a ground state rotational constant of 6.36 $\pm$ 0.03 cm$^{-1}$, in close agreement with the literature \cite{Zhu1992}. The large narrow peak corresponds to the (2+1) 3p-7p REMPI transition in aluminum \cite{Mitchell1983}. The spectrum is largely free of spectral interference, confirming that AlH is selectively photoionized. The strength of the S-branch transitions can also be used to assign a rotational temperature of 336 $\pm$ 31 K (Fig.~\ref{fig:spec}b). This temperature corresponds well with the translational temperature of 326 $\pm$ 5 K obtained from the time-of-flight profile data. The translational temperature represents an upper bound because no correction was made to the fit for the time the gas resides in the nozzle. Although substantial cooling has taken place from the molten source, the temperature is fairly high compared with what can be achieved in typical supersonic expansions. We attribute this somewhat high temperature primarily to the low conductance of our compact apparatus (50 l/s pumping speed) compared to other conventional systems (4500 l/s) \cite{Duncan2012}. Colder rotational and translational temperatures in the same apparatus could presumably also be achieved by lowering the pulse repetition rate as was performed in \cite{Tarbutt2002}. Lower temperatures can also be reached by using a smaller nozzle inner diameter, minimizing the required gas load and allowing the expansion chamber to reach lower average pressures.

\subsection{Long-Term Signal Stability}

\begin{figure}
\centering
\includegraphics[scale=0.37]{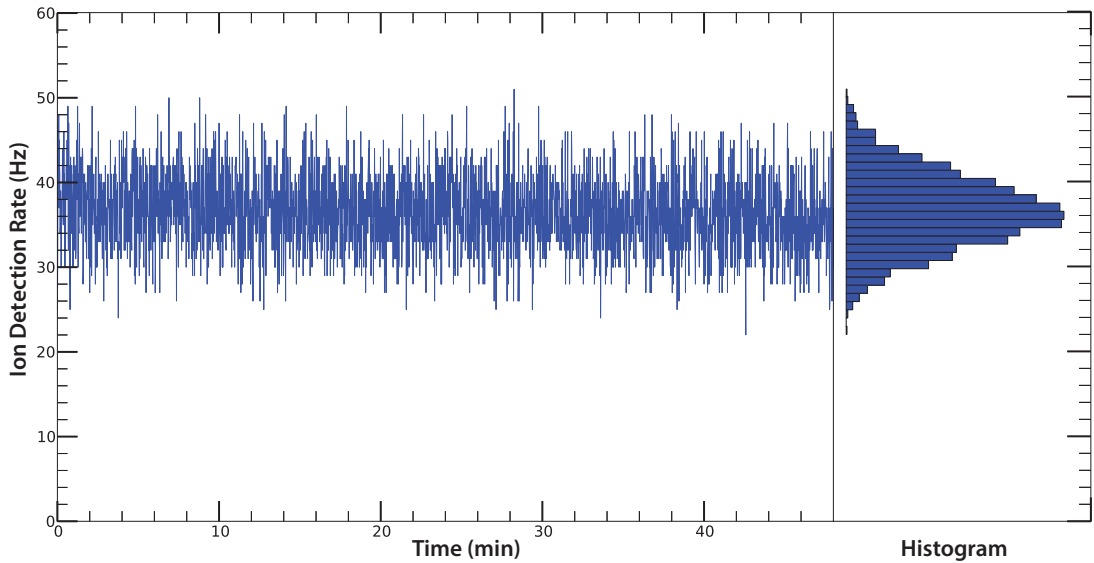}
\caption{\label{fig:stab}AlH$^+$ production rate at a constant frequency monitored over long timescales, as time series as well as associated histogram. Background ion production at this wavelength was found to be negligible.}
\end{figure}

To demonstrate the long-term stability of the source, we tune the OPO to the center of the Q branch. As seen in Fig.~\ref{fig:stab}, the ion production rate remains stable over hour timescales, also indicated by the symmetric nature of the histogram. In the first few hours of source operation, we observe a slight decrease in signal of $\sim$10\% in 3 hours. During the measurement, we also observe a gray/black film coating the inner wall of the glass nozzle. The ablation laser simultaneously back-ablates this film off of the inner nozzle wall, leaving a transparent circle for the ablation laser to reach the aluminum target \cite{Bullock1999}. After many pulses the rate of coating and back-ablation come to an equilibrium and long-term signal stability is reached. The inner nozzle wall can be cleaned off in this way by translating the ablation laser along the nozzle \footnote{Outside vacuum, a concentrated solution (2M) of potassium hydroxide (KOH) or sodium hydroxide (NaOH) is sometimes used to clean aluminum from the glass surface. The reaction releases bubbles of hydrogen gas.}. 

We also conducted measurements with a nozzle containing a 1 mm hole for the ablation laser. Over time Al coats the upper viewport. Due to the lower ablation laser intensity at the viewport (compared to at the nozzle inner wall), the viewport must be cleaned periodically with higher ablation laser power in order to recover signal. Although this nozzle design provides a long-term AlH photoionization signal, the AlH flux is more than a factor of two smaller than that produced with the closed nozzle.

\section{Conclusion}

We have demonstrated a long-term stable, pulsed beam of AlH from which AlH$^+$ can be produced via (2+1) REMPI. Using a transparent nozzle we forgo having to drill a hole for the ablation laser, allowing for increased overall beam intensity and lower internal and external temperatures. The compact size of the nozzle allows for easy integration with many UHV systems. The nozzle also accommodates using a molten Al target, providing a stable AlH flux due to the constant and smooth surface topography over many ablation pulses. We have demonstrated a 330 K AlH source and provide suggestions for reaching lower rotational and translational temperatures. Our AlH beam is directed toward an ion trap, so that AlH can be photoionized leading to a pure trapped sample of AlH$^+$. Because AlH itself is a Doppler cooling candidate \cite{DiRosa2004}, a stable beam of AlH could also potentially be used to realize laser slowing and trapping of AlH.

\ack

This work was supported by (U.S.) Air Force Office of Scientific Research (USAFOSR) FA9550-13-1-0116 and National Science Foundation (NSF) GRFP DGE-1324585. We would also like to acknowledge the Northwestern University Instrument Shop and useful conversations with Ed Grant.

\section*{References}

\bibliography{source}

\end{document}